\documentclass[aps,prl,twocolumn,preprintnumbers,superscriptaddress]{revtex4-1}

\usepackage[usenames,dvipsnames,table]{xcolor}
\usepackage{graphicx,amsmath,amssymb,amsthm,multirow,array,bm,bbm,esint}
\usepackage[mathscr]{eucal}
\usepackage[bbgreekl]{mathbbol}
\usepackage{epsf,amsfonts}
\usepackage{hyperref}

\newcommand{\beq}{\begin{equation}}
\newcommand{\eeq}{\end{equation}}

\begin{document}

\title{Scrambling in nearly thermalized states at large central charge}

\author{Kristan Jensen}
\email{kristanj@sfsu.edu}
\affiliation{Department of Physics and Astronomy, San Francisco State University, San Francisco, CA 94132, USA}

\date{\today}

\begin{abstract}
	We study $2d$ conformal field theory (CFT) at large central charge $c$ and finite temperature $T$ with heavy operators inserted at spatial infinity. The heavy operators produce a nearly thermalized steady state at an effective temperature $T_{\rm eff}\leq T$. Under some assumptions, we find an effective Schwarzian-like description of these states and, when they exist, their gravity duals. We use this description to compute the Lyapunov exponents for light operators to be $2\pi T_{\rm eff}$, so that scrambling is suppressed by the heavy insertions.
\end{abstract}

\pacs{}

\maketitle

\textit{Introduction.}~There are many measures of quantum chaos. One recently introduced by Kitaev~\cite{kitaev} is an analogue of the Lyapunov exponent for classical dynamical systems. In systems with $N\gg 1$ degrees of freedom, the out-of-time-ordered correlation function (OTOC) of two local operators $V$ and $W$ may grow as~\footnote{Strictly speaking we use a regulated version of this correlation function as in~\cite{Maldacena:2015waa} whereby the different insertions are separated in imaginary time.}
\beq
\label{E:OTOC}
	\frac{\langle V(t)W(0)V(t)W(0)\rangle_{\beta}}{\langle VV\rangle_{\beta} \langle WW\rangle_{\beta}} = 1 - \frac{1}{N^2}e^{\lambda_L(t-t_*)} + O(N^{-3})\,,
\eeq
for $t\ll t_*=O(\ln N^2)$. Here $\lambda_L$ is a quantum ``Lyapunov'' exponent, $t_*$ is the ``scrambling time'' at which the second term is $O(1)$ and before which the approximative expression above breaks down, and the correlation functions are evaluated at inverse temperature $\beta$. 

Maldacena, Shenker, and Stanford~\cite{Maldacena:2015waa} have argued that $\lambda_L$ is bounded from above as
\beq
\label{E:MSS}
	\lambda_L \leq \frac{2\pi}{\beta}\,.
\eeq
Systems which saturate the bound are sometimes said to be ``maximally chaotic.'' One reason for recent interest in maximal Lyapunov growth is that it diagnoses dual gravitational physics. CFTs with an Einstein gravity dual in Anti de Sitter (AdS) spacetime universally saturate the bound~\cite{Shenker:2013pqa}. In $2d$ CFT~\cite{Roberts:2014ifa} and in $d>2$ Rindler CFT~\cite{Maldacena:2015waa} this fact is understood to be a consequence of conformal invariance, large $N$ factorization, and a large higher spin gap. Similar statements apply for nearly conformal systems in one dimension, like the Sachdev-Ye-Kitaev (SYK) model~\cite{Sachdev:1992fk,kitaev,Polchinski:2016xgd,Jevicki:2016bwu,Maldacena:2016hyu} at large $N$ and low-temperature and Jackiw-Teitelboim gravity on nearly AdS$_2$ spacetime~\cite{Almheiri:2014cka,Jensen:2016pah,Maldacena:2016upp,Engelsoy:2016xyb}, both of which are governed by a Schwarzian effective description.

That Schwarzian model may then be viewed as an effective theory for maximal chaos. Building upon that model, as well as the Schwarzian-like description of SYK chains~\cite{Gu:2016oyy,Davison:2016ngz}, the authors have proposed a more general hydrodynamical description of maximally chaotic systems~\cite{Blake:2017ris}. A crucial feature of their model is the ``pole-skipping'' phenomenon, discovered in~\cite{Grozdanov:2017ajz}, whereby the thermal two-point function of the stress tensor exhibits a line of poles and a line of zeros as a function of momentum in the complex frequency plane, and these two cross at the frequency corresponding to the exponent $\lambda_L$ in the OTOC~\eqref{E:OTOC} (and at the momentum corresponding to the ``butterfly velocity,'' e.g.~\cite{Roberts:2016wdl}). This phenomenon arises in theories with an Einstein~\cite{Blake:2018leo} gravity dual (and persists with higher curvature corrections~\cite{Grozdanov:2018kkt}).

What of submaximal chaos in gravity or large $N$ CFT? An EFT description of submaximal chaotic growth has so far been elusive. In gravity, it is known that the exchange of a Regge trajectory of higher spin states perturbatively shifts the exponent~\cite{Shenker:2014cwa}, and there is some hope that this exchange has a ``pomeron''-like effective description.

In this note we take a different tack, and find a simple example of a large $N$ system which, in a sense, displays submaximal chaotic growth on account of stress tensor (and descendant) exchange. Consider a CFT$_2$ at large central charge $c$ on the line at finite temperature. Here $c$ plays the role of the large $N$ parameter. Insert identical  ``heavy'' operators of equal right- and left-scaling weights $h=\bar{h} = O(c)$ at spatial infinity.~\footnote{Our analysis may be easily generalized to allow for $h\neq \bar{h}$.} These insertions generate a non-equilibrium steady state~\footnote{See e.g.~\cite{Chang:2013gba,Bhaseen:2013ypa} for complementary studies of steady states in CFT$_2$.}, in which one-point functions of local operators are constant, and to leading order at large $c$ two-point functions of ``light'' operators with scaling weights $\Delta,\bar{\Delta} = O(c^0)$ are thermal at an effective temperature which depends on $h$. However the $1/c$ corrections to those two-point functions are not exactly thermal, and so the steady state is only approximately thermalized. In any case we study scrambling in the nearly-equilibrium state through an OTOC, now a six-point function with two heavy insertions. For light operators and for operators with scaling dimension $\Delta = O(\sqrt{c})$, we find the Lyapunov growth
\beq
\label{E:twistedLyapunov}
	\lambda_L =\frac{2\pi}{\beta_{\rm eff}}= \frac{2\pi}{\beta} \sqrt{1-\frac{24h}{c}}\leq \frac{2\pi}{\beta}\,,
\eeq
when $h< c/24$. This growth is ``submaximal'' with respect to the initial temperature, and ``maximal'' with respect to the effective temperature, although strictly speaking the bound~\eqref{E:MSS} does not constrain this exponent since it comes from a six-point function. We also find that the pole-skipping phenomenon of~\cite{Blake:2017ris} holds here.

To obtain these results we argue for an effective field theory (EFT) description of these systems, appropriate for CFT$_2$ in which nonlinear response is dominated by the exchange of the identity and its Virasoro descendants. This EFT may be thought of as a higher-dimensional version of the Schwarzian model. We also study the gravity dual to these CFTs when they exist. The dual geometry is the double Wick-rotation of global AdS$_3$ with a conical singularity, a two-sided BTZ black brane with a defect extended along the bifurcation surface. Starting from the global AdS$_3$ description, we can derive the EFT from gravity following~\cite{Cotler:2018zff}.

 The validity of this EFT rests on the usual properties of ``holographic'' CFTs, including large $N$ factorization and a large higher spin gap. We also note that this EFT is closely related to a variant of the Schwarzian model~\cite{Stanford:2017thb} (see also~\cite{Anninos:2018svg}) which describes submaximal Lyapunov growth~\cite{Yoon:2019cql} in some modifications of the SYK model~\cite{Ferrari:2019ogc} with twisted boundary conditions for the SYK fermions.

\emph{Note:} As this work was nearing completion a closely related paper~\cite{David:2019bmi} appeared on arXiv.

\textit{Setup.}~Consider a CFT$_2$ on the Euclidean plane with a conformally invariant vacuum. Insert identical operators $\mathcal{O}$ of scaling weights $(h,h)$ at the origin and infinity. The stress tensor acquires a nonzero one-point function
\beq
\label{E:TonePt}
	\langle T(z)\rangle_{\rm plane} = \frac{h}{z^2}\,, \qquad \langle \bar{T}(\bar{z})\rangle_{\rm plane} = \frac{h}{\bar{z}^2}\,.
\eeq
The global conformal symmetry is broken to the subgroup generated by $L_0$ and $\bar{L}_0$. 
Now conformally transform to the Euclidean cylinder of circumference $\beta$, $z = e^{2\pi i w/\beta}$. The Virasoro zero modes are $L_0^{\rm cyl} = L_0 - \frac{c}{24}$ and $\bar{L}_0^{\rm cyl} = \bar{L}_0 - \frac{\bar{c}}{24}$, with $c$ and $\bar{c}$ the right and left central charges. We take $c=\bar{c}$ in what follows. This setup admits two rather different continuations to real time. In the usual one, we take the non-compact coordinate on the cylinder to be Euclidean time, so that the real-time theory lives on a Lorentzian cylinder and the insertions are in the infinite past and future. The Euclidean path integral prepares a simultaneous eigenstate $|h,h\rangle$ of $L_0$ and $\bar{L}_0$ of the CFT on a circle of size $\beta$. Unitarity implies that $h\geq 0$, as does the averaged null energy condition on the Lorentzian plane (upon  continuing~\eqref{E:TonePt} to real time).

In the other continuation we regard the angular direction as Euclidean time, in which case we have CFT on the line at finite temperature. The insertions are now at left and right spatial infinity, and we may interpret the cylinder partition function as computing an unnormalized trace $\mathcal{Z} = \text{tr}\left(e^{-\beta H} \mathcal{O}(\infty)\mathcal{O}(-\infty)\right)$, where $H$ is the CFT Hamiltonian on the line. Assuming that the spectrum of operators has a gap above the identity, this trace has Cardy-like growth at large energy, characterized by a density of states
\beq
\label{E:rho}
	\rho(L_0,\bar{L}_0) \approx \exp\left( 2\pi\sqrt{\frac{ L_0}{6}(c-24h)} + 2\pi\sqrt{\frac{\bar{L}_0}{6}(c-24h)} \right)\,,
\eeq
provided that $c>24h$.~\footnote{The same expression holds if the vacuum is not conformally invariant, but instead has ``vacuum charges'' $h=\bar{h}$. In that case the Cardy-like growth~\eqref{E:rho} is sometimes~\cite{Carlip:1998qw} phrased as the statement that there is an effective central charge $c_{\rm eff} = c - 24h$.} Equivalently, there is a ``free energy density'' $\mathcal{Z} = e^{-\beta V f}$ with $V$ the regularized spatial volume,
\beq
\label{E:twistedF}
	f \approx - \frac{\pi \,c}{6\beta^2}\alpha^2\,, \quad \alpha = \sqrt{1-\frac{24 h}{c}}\,.
\eeq
In what follows we take $h = O(c)$ and $h<\frac{c}{24}$, for which the expressions above are valid. Note that $f$ takes the same form as the thermal free energy density of a CFT$_2$ on the line, at an effective temperature $\frac{1}{\beta_{\rm eff}} = \frac{1}{\beta}\sqrt{1-\frac{24 h}{c}} \leq \frac{1}{\beta}$.

As this result suggests, the thermal state perturbed by the insertions at infinity is approximately thermal at a temperature $1/\beta_{\rm eff}$. In fact it is a \emph{non-equilibrium steady state}. To wit, suppose we probe this state with local operators. It is simple to show that the one-point function of the stress tensor takes the usual thermal form at temperature $1/\beta_{\rm eff}$. More generally one-point functions of local Virasoro primaries $V$ of scaling weights $(\Delta,\bar{\Delta})$ are constants, proportional to the $\mathcal{O}\mathcal{O}V$ operator product expansion coefficient $C_{\mathcal{O}\mathcal{O}V}$,
\beq
	\langle V\rangle_{\beta,h} = \frac{\text{tr}(e^{-\beta H}\mathcal{O}(\infty)\mathcal{O}(-\infty)V)}{\text{tr}(e^{-\beta H}\mathcal{O}(\infty)\mathcal{O}(-\infty))} = C_{\mathcal{O}\mathcal{O}V} \left( \frac{2\pi}{\beta}\right)^{\Delta + \bar{\Delta}}\,.
\eeq
So we have a steady state. This does not quite match thermal physics, as the thermal one-point functions of primaries vanish. However at large $c$ it is often the case that the $C$'s are suppressed, in which case the one-point functions are approximately thermal.

What of higher-point functions? $n$-point functions of local operators in the state may be obtained from $n+2$-point functions on the plane, where the additional two insertions are the $\mathcal{O}$'s. For example, the two-point function of two operators $V$ in the state comes from a four-point function $\langle \mathcal{O}VV\mathcal{O}\rangle_{\rm plane}$. In this work we assume not only large $c$ but ``vacuum dominance,'' meaning that we study CFT$_2$ for which the dominant contribution to these correlation functions arises from the exchange of the identity operator and its Virasoro descendants in some channel. (For a higher spin gap of $O(c)$, this approximation will necessarily break down at cross-ratios of order $1/c$.) Thus, to understood two-point or four-point functions in the state we require multi-point Virasoro identity blocks with two heavy insertions. As we will see, known results for the blocks at large $c$ with an an $O(1)$ number of light insertions imply that the state is approximately thermal.

Now suppose that the CFT$_2$ has an Einstein gravity dual, with $c=\bar{c} \approx \frac{3}{2G}\gg 1$ and we normalize the AdS radius to unity. For $h=O(c)$ with $h<\frac{c}{24}$ the dual gravitational background to the CFT on the Euclidean cylinder is
\begin{align}
\begin{split}
\label{E:euclideanBTZ}
	ds^2 &= \frac{4 dz d\bar{z} + r_h^2(1+|z|^2)^2dy^2}{(1-|z|^2)^2}\,,
	\\
	z &= r e^{2\pi i\alpha  \tau/\beta}\,, \quad r_h = \frac{2\pi}{\beta}\alpha\,, 
\end{split}
\end{align}
where $\tau\sim \tau+ \beta$, and $y$ is non-compact. There is a defect at $r=0$, which produces a deficit angle $2\pi(1-\alpha)$. This defect is smoothed out at finite $c$. The free energy density of the black brane is $f \approx -\frac{\pi c}{6\beta^2}\alpha^2$, matching~\eqref{E:twistedF}, and the entropy density is $s \approx \frac{\pi c}{3\beta}\alpha^2$. 

There are two possible continuations to real time. The first, in which $y$ is Euclidean time, leads to global AdS$_3$ with a conical singularity. The second, corresponding to $\tau$ being Euclidean time leads to a two-sided black brane via $z \to u = x+t, \bar{z}\to -v = x-t$, 
\beq
\label{E:kruskal}
	ds^2 = \frac{-4du dv + r_h^2(1-uv)^2dy^2}{(1+uv)^2}\,,
\eeq
however with a defect at the bifurcation surface $u=v=0$. The boundary conditions for bulk fields at the horizons $u=0$ and $v=0$ are also modified. To understand those boundary conditions we perform a Hartle-Hawking construction of the bulk state as in~\cite{Maldacena:2001kr}. The state is prepared by a Euclidean section, half of the Euclidean geometry~\eqref{E:euclideanBTZ}, covered by $\tau \in [0,\beta/2]$. To evolve this state we glue the boundary of this section, at $\tau = 0,\beta/2$, to the future half of the Lorentzian geometry~\eqref{E:kruskal} at the $t=0$ slice. The bifurcation surface is glued to the conical deficit at the center of the Euclidean section. In the limit that the conical defect is infinitely sharp, the stress tensor for matter fields is singular there, and so also at the horizons of the Lorentzian section. This singularity is regulated at finite $c$.

We are interested in the boundary stress tensor, i.e. the fluctuations of the gravitational field on top of this background. These fluctuations are well-behaved in the presence of the conical singularity owing to the topological nature of pure three-dimensional gravity, and so for us the precise resolution of the singularity is unimportant.

\begin{figure}[t]
\includegraphics[width=2.5in]{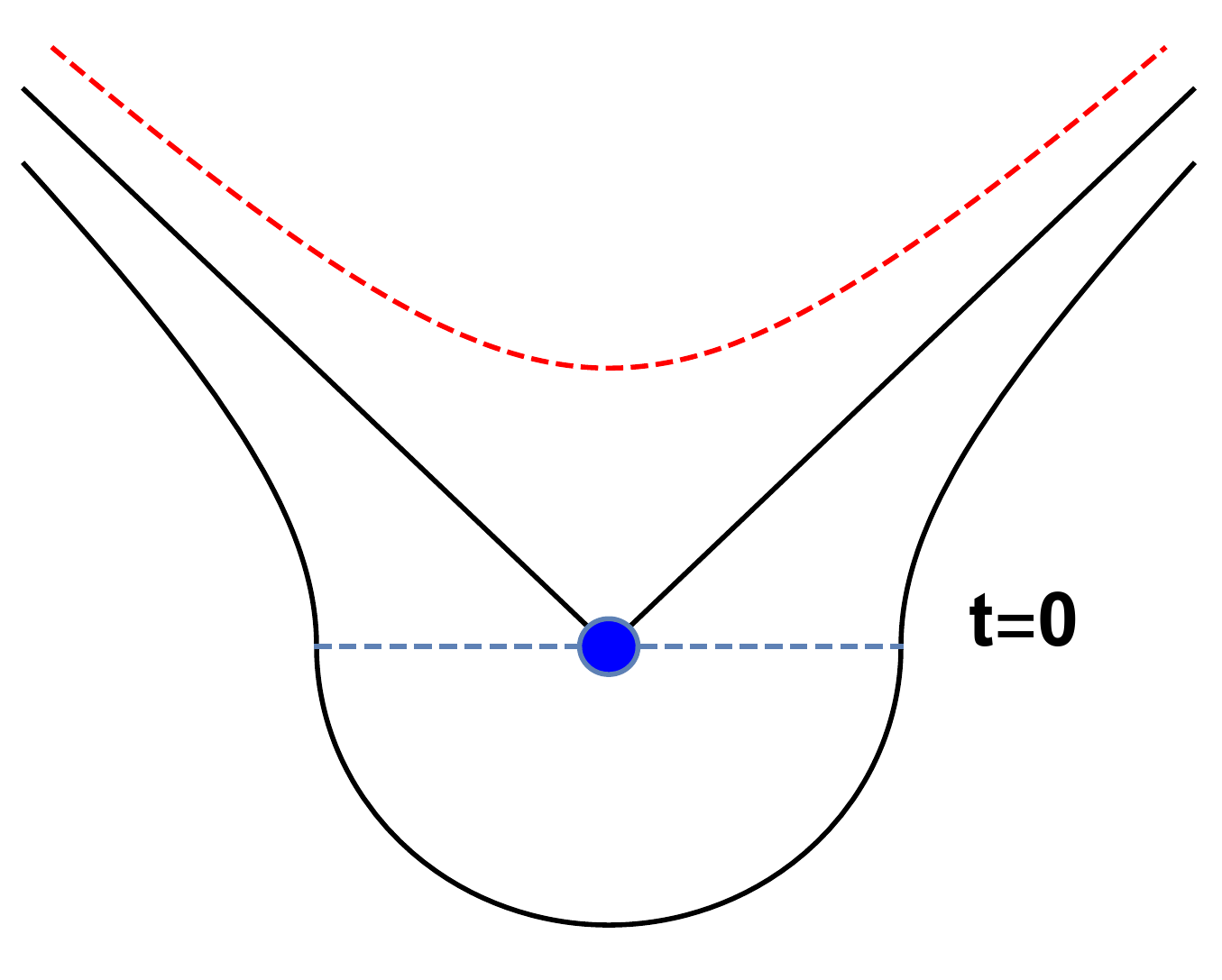}
\caption{\label{F:HH} The Hartle-Hawking construction for the two-sided BTZ black brane with nonzero $h, \bar{h}$. Half of the Euclidean geometry~\eqref{E:euclideanBTZ} is glued to the future half of the two-sided black brane across the $t=0$ slice. The thick blue dot represents the defect, the diagonal black lines the horizons, and the dashed red line the future singularity.}
\end{figure}

\textit{EFT and pole-skipping.}~We would like to compute correlation functions assuming vacuum dominance, i.e. Virasoro identity blocks with two heavy insertions. This may be done at large $c$ via standard techniques~\cite{Fitzpatrick:2015zha}. See e.g.~\cite{Banerjee:2016qca,Gao:2018yzk,Anous:2019yku}. In fact, the six-point block with two heavy insertions and four insertions which are either light or ``hefty''~\footnote{This moniker is due to E.~Dyer.} has been computed in~\cite{Anous:2019yku}. By ``hefty'' we mean that their scaling weights are of $O(\sqrt{c})$. Those authors use the block for a different goal, to find the Lyapunov exponent for light operators in an eigenstate prepared by a heavy operator with $h>\frac{c}{24}$, which has an effective temperature set by $h$. They find $\lambda_L = \frac{2\pi}{\beta_{\rm eff}}$. For our purposes we reinterpret the block as a twisted thermal four-point function of light operators, from which we can obtain the OTOC~\eqref{E:OTOC} in the steady state by an appropriate continuation to real time. This gives the Lyapunov exponent~\eqref{E:twistedLyapunov} advertised above. 

We develop an EFT approach to computing the blocks along the lines of~\cite{Cotler:2018zff}, independently computing the six-point block with four light or ``hefty'' insertions. The result agrees with that of~\cite{Anous:2019yku}.

The starting point is to consider the continuation where we have the CFT on the cylinder in the state $|h,h\rangle$. The Virasoro symmetry is spontaneously broken and the relevant coset manifold is two copies of Diff$(\mathbb{S}^1)/U(1)$, infinitesimally parametrized by the $L_{n\neq 0}$ and $\bar{L}_{m\neq 0}$. There is a simple effective theory which realizes this symmetry breaking pattern, and which matches the anomalies of the underlying CFT: two copies of the Alekseev-Shatashvili path integral quantization~\cite{Alekseev:1988ce} of the space Diff$(\mathbb{S}^1)/U(1)$, a coadjoint orbit of the Virasoro group. See~\cite{Cotler:2018zff} for an extensive discussion of how to perform this path integral. The model depends on two independent, real scalar fields $\phi$ and $\bar{\phi}$ which obey the boundary condition and identification
\beq
	\phi(\theta+2\pi,t) = \phi(\theta,t)+2\pi\,, \quad \phi(\theta,t) \sim \phi(\theta,t)+a(t)\,,
\eeq
and similarly for $\bar{\phi}$. That is, at fixed time, $\phi$ and $\bar{\phi}$ are elements of the quotient space Diff$(\mathbb{S}^1)/U(1)$. The effective action is
\begin{align}
\begin{split}
\label{E:EFT}
	S =& -\frac{C}{24\pi}\int d^2x \left( \frac{\phi''\partial_+\phi'}{\phi'^2}-\alpha_0^2\phi'\partial_+\phi\right)
	\\
	&\quad -\frac{C}{24\pi} \int d^2x \left( \frac{\bar{\phi}''\partial_-\bar{\phi}'}{\bar{\phi}'^2} - \alpha_0^2 \bar{\phi}'\partial_-\bar{\phi}\right)\,,
\end{split}
\end{align}
where $c=C+1$ is the central charge~\cite{Cotler:2018zff} and $\alpha_0 = \sqrt{1 - \frac{24 h}{C}}$. Further $' = \partial_{\theta}$ and $x^{\pm} = \theta \pm t$ are null coordinates on the cylinder. There is a unique classical saddle modulo the identification, $\phi_0 = \bar{\phi}_0 = \theta$, and canonically normalizing fluctuations around the saddle, one finds that the theory is weakly coupled at large $C$.

This model may also be derived from AdS$_3$ gravity with $C = \frac{3}{2G}$. Analytically continuing $y$ to real time, the Euclidean geometry~\eqref{E:euclideanBTZ}  becomes Lorentzian global AdS$_3$ with a defect at its center. Using the Chern-Simons formulation~\cite{Achucarro:1987vz,Witten:1988hc} of pure three-dimensional gravity, in terms of which the defect is generated by a Wilson line, one can rewrite the gravitational path integral in this setting as a constrained chiral Wess-Zumino-Witten model on the AdS boundary. This was done in~\cite{Cotler:2018zff}, with the result above.

To go from here to steady state correlation functions we perform the double Wick rotation,
\beq
	t \to -i \frac{2\pi y}{\beta}\,,\qquad  \theta = \frac{2\pi \tau}{\beta}\,, \qquad \tau \to i t\,,
\eeq
with an appropriate $i\epsilon$ prescription. We work in the Euclidean theory, and then continue to real time at the end.

To investigate pole-skipping we require the retarded thermal two-point function of the stress tensor. It may be deduced using the Virasoro algebra and the operator/state correspondence, but we find it instructive to compute it using our EFT. Let us focus on the holomorphic sector. The holomorphic stress tensor is
\beq
	T= -\frac{C}{24}\left( \{\phi,\theta\} + \frac{\alpha_0^2}{2}\phi'^2\right)\,,
\eeq
where $\{f(x),x\} =\frac{f'''(x)}{f'(x)} - \frac{3}{2} \frac{f''(x)^2}{f'(x)^2}$ is the Schwarzian derivative, and similarly for $\bar{T}$. Expanding around the classical saddle, $\phi = \frac{2\pi\tau}{\beta} + \varepsilon$, the stress tensor has a linear coupling to $\varepsilon$
\beq
\label{E:linearstress}
	T = -\frac{C\alpha_0^2}{48} - \frac{C}{24}(\partial_{\theta}^2 + \alpha_0^2)\partial_{\theta} \varepsilon + O(\varepsilon^2)\,.
\eeq
The quadratic action for $\varepsilon$ obtained from~\eqref{E:EFT} is
\beq
	S_2 = \frac{ C}{24\beta}\int d\tau dy \,\varepsilon(\partial_{\theta}^2 + \alpha_0^2) (\partial_{\tau}+i \partial_y)\varepsilon\,,
\eeq
from which we obtain the $\varepsilon$ propagator in Fourier space
\beq
\label{E:propagator}
	\langle \varepsilon(n,k)\varepsilon(-n,-k)\rangle = \frac{6 \beta^2\alpha_0^2}{\pi C}\frac{1}{i n(n^2-\alpha_0^2)(\frac{\beta k}{2\pi}-i n)}\,,
\eeq
where $n$ labels Matsubara frequencies $\omega_n = \frac{2\pi n}{\beta}$ around the $\tau$-circle and $k$ is $y-$momentum. The connected two-point function of $T$ at large $c$ comes from a single exchange of $\varepsilon$. Computing this diagram and continuing to real time, we extract the connected, retarded two-point function of $T_{++}$ in Fourier space (where now $x^{\pm} = y \pm t$),
\beq
	G^R_{++,++}(\omega,k) \approx \frac{ c \,\alpha^2}{96\pi}\frac{\omega\left(\omega^2+\left(\frac{2\pi \alpha}{\beta}\right)^2\right)}{\omega +  k+ i \epsilon}\,.
\eeq
Up to a change in normalization, this is the usual thermal two-point function of a CFT$_2$~\cite{Haehl:2018izb} on the line at a rescaled temperature $\frac{1}{\beta} \to \frac{\alpha}{\beta}$. (Indeed, using the methods of~\cite{Haehl:2018izb} one could reconstruct the quadratic approximation to this EFT.) There is a line of poles at $\omega = -k$, and zeros at $\omega=0, \pm \frac{2\pi i}{\beta}\alpha$. The lines cross and so poles are skipped at $\omega =k = 0$ and at $\omega = -k=\pm \frac{2\pi i}{\beta}{\alpha}.$ The $+$ branch corresponds to an exponentially growing mode $e^{\frac{2\pi}{\beta}\alpha(t+y)}$. As we see below this mode yields the Lyapunov exponent and a butterfly velocity equal to the speed of light.

\textit{Correlators.}~We move on to consider correlation functions of local operators in the steady state. In addition to local operators like the stress tensor, the EFT has bilocal operators, reparameterized two-point functions of dimension $\Delta$ primaries at the same value of $y$,
\beq
\label{E:bilocal}
	\mathcal{B}(\tau_1,\tau_2,y,;\Delta) =\left( \frac{\partial_{\tau_1}f(\tau_1,y)'\partial_{\tau_2}f(\tau_2,y)}{(f(\tau_1,y)-f(\tau_2,y))^2}\right)^{\Delta}\,,
\eeq
with $f(\tau,y) = \tan\left( \frac{\alpha_0 \phi(\tau,y)}{2}\right)$. There is substantial evidence~\cite{Cotler:2018zff} that the correlation functions of these bilocal operators in the EFT above encode holomorphic Virasoro identity blocks. Effectively, the reparameterization that appears in~\cite{Fitzpatrick:2015zha} is the dynamical degree of freedom $\phi$. In the Chern-Simons formulation of pure $3d$ gravity, the bilocal is a boundary-anchored Wilson line (see e.g.~\cite{Bhatta:2016hpz,Besken:2016ooo,Fitzpatrick:2016mtp} for works relating these Wilson lines to Virasoro blocks).

Consider a light operator $V$ of dimension $\Delta = \Delta_V$. To leading order at large $c$ the one-point function of $\mathcal{B}$ is obtained by plugging the classical trajectory $\phi = \frac{2\pi \tau}{\beta}$ into $\mathcal{B}$, giving
\beq
	\langle \mathcal{B}(\tau_1,\tau_2,y;\Delta_V\rangle = \left( \frac{\pi \alpha}{\beta \sin\left( \frac{\pi \alpha \tau_{12}}{\beta}\right)}\right)^{2\Delta_V}\left( 1 + O(c^{-1})\right)\,,
\eeq
which, after appropriate translation, is the leading order result for the heavy-heavy-light-light (HHLL) block obtained in~\cite{Fitzpatrick:2015zha}. The $1/c$ correction to the block was computed in~\cite{Fitzpatrick:2015dlt}, and the $1/c$ correction to $\langle \mathcal{B}\rangle$ was computed in~\cite{Cotler:2018zff}; the two agree. So the EFT computes the HHLL block. Approximating the two-point function of $V$ by the block and continuing to real time, we have
\beq
	\langle V(t,y)V(0,0)\rangle_{\beta,h} \approx \left( \frac{\pi \alpha}{\beta \sinh\left( \frac{\pi \alpha}{\beta}(t+y)\right)}\right)^{2\Delta_V}\,,
\eeq
which is the usual form for a thermal two-point function in CFT$_2$ at the same effective temperature as before $\frac{\alpha}{\beta}$. However, the $1/c$ correction to the block is not periodic under $\tau \to \tau + \frac{1}{\beta_{\rm eff}}$. So the two-point function of $V$ is only approximately thermal at large $c$. This is the sense in which the steady state is \emph{nearly} thermalized.

Now consider the $2n$-point function of two light $V_1$'s, two light $V_2$'s, and so on. This is approximated by a block with two heavy insertions and $2n$ light insertions, which we compute in our theory through the $n$-point function of bilocals. At large $c$ this $n$-point function factorizes
\begin{align}
	\langle &\mathcal{B}_{1,2}(\Delta_{V_1})\hdots \mathcal{B}_{2n-1,2n}(\Delta_{V_n})\rangle 
	\\
	\nonumber
	&=\langle \mathcal{B}_{1,2}(\Delta_{V_1})\rangle \hdots \langle \mathcal{B}_{2n-1,2n}(\Delta_{V_n})\rangle \left( 1 + O(c^{-1})\right)\,,
\end{align}
which recovers the known result for the HHL$\hdots$L block~\cite{Banerjee:2016qca}. So $2n$-point functions of light operators in the steady state factorize into products of approximately thermal two-point functions at temperature $1/\beta_{\rm eff}$.

Now we wish to evaluate an OTOC in the steady state. Consider the normalized four-point function of two light operators $V$ and two light operators $W$,
\beq
	\mathcal{F}(\tau_i,y_i) = \frac{\langle V(\tau_1,y_1)V(\tau_2,y_2) W(\tau_3,y_3)W(\tau_4,y_4)\rangle_{\beta,h}}{\langle V(\tau_1,y_1)V(\tau_2,y_1)\rangle_{\beta,h}\langle W(\tau_3,y_3)W(\tau_4,y_4)\rangle_{\beta,h}}\,.
\eeq
Approximating it as the identity block, we compute it in our EFT at $y_1=y_2=y$, $y_3=y_4=0$ via
\beq
	\mathcal{F} =  \frac{\langle \mathcal{B}(\tau_1,\tau_2,y;\Delta_V)\mathcal{B}(\tau_3,\tau_4,0;\Delta_W)\rangle}{\langle \mathcal{B}(\tau_1,\tau_2,y;\Delta_V)\rangle\langle\mathcal{B}(\tau_3,\tau_4,0;\Delta_W)\rangle}\,,
\eeq
At leading order in large $c$ this two-point function factorizes, $\mathcal{F} = 1 + O(c^{-1})$. The bilocal has a linear coupling
\begin{align}
\nonumber
	\mathcal{B}_{1,2}(\Delta) &= \left( \frac{\pi \alpha_0}{\beta\sin\left( \frac{\pi \alpha_0 \tau_{12}}{\beta}\right)}\right)^{2\Delta}\left(1 +\Delta \mathcal{J}_{12}\cdot \varepsilon + O(\varepsilon^2)\right)\,,
	\\
	\mathcal{J}_{12}\cdot \varepsilon &= \varepsilon_1'+\varepsilon_2' -  \alpha_0\cot\left( \frac{\pi \alpha_0 \tau_{12}}{\beta}\right)\epsilon_{12}\,,
\end{align}
where $'$ indicates a $\theta$-derivative. The $O(c^{-1})$ term in $\mathcal{F}$ is computed by a diagram with a single exchange. In terms of the linear coupling we have $\mathcal{F} = 1 + \Delta_V \Delta_W \langle (\mathcal{J}_{12}\cdot\varepsilon)(\mathcal{J}_{34}\cdot\varepsilon)\rangle + O(c^{-2})$. Fourier transforming the $\varepsilon$ propagator~\eqref{E:propagator} to configuration space, and holomorphically continuing to general $y_i$ we find after some manipulation
\beq
	\mathcal{F} = 1+  \frac{2\Delta_V \Delta_W}{c}u^2\,_2F_1(2,2;4;u) + O(c^{-2})\,,
\eeq
where
\beq
	u = \frac{(\zeta_1^{\alpha} - \zeta_2^{\alpha})(\zeta_3^{\alpha}-\zeta_4^{\alpha})}{(\zeta_1^{\alpha}-\zeta_3^{\alpha})(\zeta_2^{\alpha}-\zeta_4^{\alpha})}\,, \qquad \zeta_i = e^{\frac{2\pi i (\tau_i+i y_i)}{\beta}}\,.
\eeq
The correction is a reparameterized global conformal block encoding the exchange of the stress tensor, and it matches the result of~\cite{Anous:2019yku} for the six-point block. Note that the only effect of the heavy insertions is to rescale the temperature to be $1/\beta_{\rm eff}$. Indeed, performing a second sheet continuation as in~\cite{Roberts:2014ifa} to extract the OTOC, we find an exponential growth $\sim \frac{\Delta_V \Delta_W}{c} e^{\frac{2\pi \alpha}{\beta}(t+y)}$, so that the Lyapunov exponent is~\eqref{E:twistedLyapunov} with a butterfly velocity equal to the speed of light. This exponentially increasing mode coincides with skipped pole in the retarded two-point function of the stress tensor.

We may also extend our analysis to obtain the six-point block with four ``hefty'' insertions with $\Delta = O(\sqrt{c})$. It is already known that the global stress tensor block exponentiates in the absence of heavy insertions~\cite{Fitzpatrick:2014vua}. The same holds true with the heavy insertions. To see this, we use that upon rescaling $\varepsilon \to \varepsilon/\sqrt{C}$ so that its propagator is $O(c^0)$ and taking $\Delta = \mathfrak{h} \sqrt{C}$ with $\mathfrak{h}$ fixed, the linear coupling of the bilocal exponentiates,
\beq
	\mathcal{B}_{1,2}(\Delta) = \left( \frac{\pi \alpha_0}{\beta \sin\left( \frac{\pi \alpha_0 \tau_{12}}{\beta}\right)}\right)^{2\Delta}e^{\mathfrak{h} \mathcal{J}_{12}\cdot \varepsilon}\left( 1 + O(1/\sqrt{c})\right)\,.
\eeq
Then $\mathcal{F}$ becomes
\beq
	\mathcal{F} = \langle \langle e^{\mathfrak{h}_V \mathcal{J}_{12}\cdot\varepsilon} e^{\mathfrak{h}_W \mathcal{J}_{34}\cdot \varepsilon}\rangle\rangle \left( 1 + O(1/\sqrt{c})\right)\,,
\eeq
which exponentiates the single exchange,
\beq
	\mathcal{F} = \exp\left( 2 \mathfrak{h}_V \mathfrak{h}_W u^2 \,_2F_1(2,2;4;u)\right)\left( 1 + O(1/\sqrt{c})\right)\,.
\eeq
As above the only effect of the heavy insertions is to replace the temperature with the effective one $1/\beta_{\rm eff}$. Consequently one finds the same Lyapunov growth as for light operators.

\textit{Acknowledgements.}~We would like to thank J.~Cotler for useful discussions. K.~J. is supported in part by the Department of Energy under grant number DE-SC 0013682. 

\bibliography{refs}

\end{document}